# 3. Hybrid codes (massless electron fluid)

D. Winske[1], Homa Karimabadi[2,3], Ari Le[4], N. Omidi[5], Vadim Roytershteyn[6], Adam Stanier[7]

[1]Los Alamos National Laboratory, Los Alamos NM, winske@lanl.gov
[2]Analytics Ventures Inc., San Diego, CA 92121, homa@analyticsventures.com
[3]AlphaTrAI, San Diego, CA 92121, homa@alphatrai.com
[4]Los Alamos National Laboratory, Los Alamos NM, arile@lanl.gov
[5]Solana Scientific Inc., Solana Beach CA, omidi@roadrunner.com
[6]Space Science Institute, Boulder CO, vroytersh@gmail.com
[7]Los Alamos National Laboratory, Los Alamos NM, stanier@lanl.gov

**Abstract** Hybrid codes are widely used to model ion-scale phenomena in space plasmas. Hybrid codes differ from full particle (PIC) codes in that the electrons are modeled as a fluid that is usually assumed to be massless, while the electric field is not advanced in time, but instead calculated at the new time level from the advanced ion quantities and the magnetic field. In this chapter we concentrate on such hybrid models with massless electrons, beginning with a discussion of the basics of a simple hybrid code algorithm. We then show examples of recent use of hybrid codes for large-scale space plasma simulations of structures formed at planetary bow shock – foreshock systems, magnetic reconnection at the magnetopause, and complex phenomena in the magnetosheath due to the interaction of kinetic processes associated with the bow shock, magnetic reconnection, and turbulence. A discussion then follows of a number of other hybrid codes based on different algorithms that are presently in active use to investigate a variety of plasma processes in space as well as some recent work on the development of new models.  We conclude with a few brief comments concerning the future development and use of hybrid codes.

# 1 Introduction

A hybrid code in the context of plasma physics generally refers to a computational model in which some of the plasma species are treated kinetically and others as fluids, while the electric and magnetic fields can be considered as either electrostatic or electromagnetic. Most often in hybrid codes employed in space plasmas all the different ion species are treated kinetically using particle-in-cell methods, the electrons are modeled as a fluid with massless electrons, and Maxwell's equations are solved in the radiation-free (i.e., no light waves) limit. It is in this more restricted domain that we discuss hybrid codes in this chapter. Even in this limited space, there is still freedom in how to treat the fast phenomena associated with electron dynamics. In space plasmas in particular, hybrid codes occupy a unique niche, intermediate between fluid codes that are most useful for modeling large portions of the solar wind or the interaction of the solar wind with the Earth's magnetosphere and fully kinetic treatment of all the plasma components, i.e., all ion species and electrons, that are resolved down to the smallest electron scales. Fully kinetic codes are commonly referred to as PIC codes, because traditionally particle-in-cell methods have been used to treat the particle dynamics. Instead, hybrid codes model a portion of the solar wind-magnetosphere boundary or a small region in the solar wind or the magnetosphere on spatial and temporal scales on which ion dynamics dominates. The effects of the larger fluid scales can be modeled through boundary conditions, while the small-scale effects of electrons are added in through models for the electron pressure tensor, and high-frequency (e.g., electron-ion lower-hybrid or higher electron-electron) wave effects are included through a resistivity or other transport coefficients. But as computers have become larger and faster, fully kinetic codes can now model system sizes and time scales that previously were only accessible to hybrid codes, and hybrid codes in turn can now model systems large enough that could once only be studied by fluid models. Thus, as the domains of calculations with different physics models can overlap, having an assortment of hybrid models allows comparisons between hybrid-PIC and fluid-hybrid simulations to be done effectively in order to verify the validity of models, provide additional validation of codes, and give further insight into spacecraft data.

This chapter is an updated version of the chapter "Hybrid Simulations Codes: Past, Present and Future – A Tutorial" in the 2003 edition of the book [1]. As we shall see, computational plasma physics has advanced so rapidly in recent years that what we called "The Future" in 2003 is now not even in the "The Present", but already fading into "The Past". However, some things have remained the same – namely the basic algorithms used in hybrid codes. On the other hand, new advances in algorithm development show excellent promise for future major advances in understanding of space physics phenomena. And present-day simulations have far exceeded what was thought to be possible 20 years ago.

In contemporary terms, the 2003 hybrid code tutorial had both strong and weak points. The old tutorial was very strong in its rather complete discussion of the assumptions and underlying equations and numerical methods of solution of the equations of the computational model as well as an extensive list of references. There was also a detailed historical tracing of the development of hybrid codes and associated numerical methods, particularly in relation to the $6^{th}$ International Space Simulation School (ISSS-6) as well as to the earlier schools. The discussion included the most common implementations of the hybrid model in simulation codes as well as examples of results from calculations and comparative test cases. Most of the material in this tutorial is still relevant today and does not need to be repeated here. In addition, issues of great interest in the time frame of ISSS-6 were considered in some detail. For example, the question of how to compute the complete electron pressure tensor and ways to include it in a generalized hybrid model for application to magnetic reconnection were well described. Also, initial attempts at using hybrid codes to model global interactions of the solar wind with planetary magnetospheres were demonstrated.

However, the old tutorial suffers now from being rather out of date concerning what is presently considered state-of-the-art. For example, the goals of performing efficient calculations on parallel computers with 1000's of processors and analyzing data from multi-satellite missions, such as CLUSTER and THEMIS, were thought at the time to be perhaps a decade in the future. In fact, however, this was achieved much sooner and present-day simulations have extended far beyond that frontier. PIC and hybrid codes are now at the threshold of running in three spatial dimensions for millions of processor hours using trillions of particles, billions of cells, and almost a million time-steps. What were then very difficult issues related to trying to embed hybrid or kinetic models in larger-scale fluid models in order to model micro-scale processes occurring in magnetic reconnection have now been replaced by very large-scale PIC codes that include all the smallest scales naturally in problems approaching global scales with realistic ratios of the ion to electron mass.

In addition to these very massive PIC and hybrid codes running on what is headed toward exa-scale computing, the computational environment is also changing in other ways. With larger computational domains, longer times, and more realistic initial/boundary conditions, the emphasis is now on investigating the interactions between various physical processes, rather than focusing on a single physics issue. This has been motivated in part by the existence of better spacecraft data, involving very high-resolution simultaneous data from multiple satellites. One example is the MMS (Magnetosphere Multi-Scale) mission [2] whose four closely-spaced satellites have fully resolved ion and have begun to explore electron scales related to magnetic reconnection layers. Better diagnostics in the simulations have led to much clearer comparisons and analysis of the spacecraft observations as well as improved insight into the underlying physical processes. With larger and longer duration calculations, the need for more sophisticated hybrid algorithms that pay careful attention to energy conservation and suppression of slow-growing numerical instabilities has become a major issue, as we will discuss later. In addition, extended hybrid models that include electron inertia effects, new quasi-neutral models, and better Vlasov methods have become more prevalent, which allow investigating short-scale electron processes that are now becoming accessible in space and laboratory observations of magnetic reconnection. We touch only briefly on these topics later, which are treated in much more detail in chapters 5, 9, and 10.

This updated chapter on hybrid codes has several purposes. The first purpose is to serve as a tutorial for newcomers to the field of computational space plasma physics by providing a brief introduction to hybrid code methodology. It includes a review of the basics of the hybrid physics model, the underlying fundamental equations that are used, and a simple example of the physics implementation into a working algorithm. This material in Sec. 2 is relatively brief, but we emphasize in the discussion that there are old as well as recent articles that cover this background material quite completely and provide updated references. It also raises the major issue inherent in all hybrid algorithms, namely how to consistently update the electric field. This issue will also appear prominently later, in Sec. 4 and Sec. 5. The purpose of Sec. 3 is to provide some contemporary examples of hybrid code use in space physics and their relation to observations. This usage is illustrated in only a few examples, but appropriate references to these and other works are presented here as well as throughout the rest of the chapter. Sec. 4 addresses other hybrid algorithms and briefly describes their use for a wide range of applications at present. Sec. 5 discusses several new hybrid algorithm development efforts, the motivation for constructing them, and their potential applications as well as appropriate numerical test cases of their use. The inclusion of electron inertia effects in advanced hybrid algorithms is also briefly summarized. Finally in Sec. 6, we look out into the future with some trepidation, recalling from the earlier version of this tutorial how rapidly and dynamically computational plasma physics can change, to suggest potential areas of growth, development, and use for hybrid codes in space and astrophysical plasmas.

## 2 Review of basic model and implementation

Hybrid codes model plasma and low-frequency electromagnetic wave phenomena on spatial and temporal scales shorter than those used in a magnetohydrodynamic description, but longer than those that resolve electron dynamics (i.e., effects occurring on electron gyroradius and electron inertial length scales and inverse electron gyrofrequency time scales). The relevant scales are then the ion gyroradius and ion inertial length and inverse ion gyrofrequency scales. Higher frequency ion physics (i.e., ion plasma frequency) that occurs on ion plasma Debye lengths is also excluded, although there are electrostatic hybrid codes that do model such processes. In space plasmas occurring within the Earth's magnetosphere and in the upstream solar wind, the relevant ion scale lengths are on the order of 10 to ~ 100 km and time scales on the order of seconds. Such time and length scales have been readily resolved by spacecraft over the past 40 years, starting with the ISEE-1,2 spacecraft that first resolved the structure of the Earth's bow shock [3]. Subsequent missions led to the wide use of hybrid codes in interpreting observations and comparing with theoretical models that will become evident in Sec. 4. Modern space instrumentation, such as found on MMS, is now able to resolve plasma distributions on much faster time scales associated with some electron phenomena and well resolve the corresponding ion scales. In fact, the four MMS spacecraft have reduced their separation to distances as small as 10 km, where electron dynamics start to become important, and have been able to make measurements down in the magnetic reconnection region [4].

In this section we present a simplified hybrid model with massless electrons, which was presented in the earlier tutorial (Winske et al. [1]), which is still commonly used, that provides a framework for discussion of more complex adaptations later in the chapter. The basic physics model and the associated equations that are solved as a function of space and time are defined as follows. To be consistent with the hybrid model, the ions are treated kinetically using standard particle-in-cell methods [5]. Each simulation ion (with position $\mathbf{x_p}$, velocity $\mathbf{v_p}$, charge $q_i$ and mass $m_i$), which represents many actual physical particles, is subject to the usual (non-relativistic) equations of motion in the electric $\mathbf{E}$ and magnetic $\mathbf{B}$ fields:

$$m_i d\mathbf{v}_p / dt = q_i (\mathbf{E} + \mathbf{v}_p \times \mathbf{B}) \, , \tag{1}$$

$$d\mathbf{x}_p / dt = \mathbf{v}_p \, , \tag{2}$$

where the fields have values given on a spatial grid, and are interpolated to the location of the simulation particle. The updated particle data for the ion charge density ($q_i n_i$) and current $\mathbf{J}_i = q_i n_i \mathbf{V}_i$, are collected on the grid, from which the number density ($n_i$) and flow velocity ($\mathbf{V}_i$) are determined for use later in calculating the electric field. (For simplicity, we assume a single ion species; for multiple species, one accumulates the data for each species separately and then adds them together to get the total ion charge density and current.)

In the standard hybrid model the electrons are treated as an massless fluid, although "inertia-less" is a more accurate description. Thus, the left side of the electron momentum equation is zero, resulting in:

$$d(n_e m_e \mathbf{V}_e)/dt = 0 = -en_e(\mathbf{E} + \mathbf{V_e} \times \mathbf{B}) - \nabla \cdot \mathbf{P_e} + en_e \eta \cdot \mathbf{J} \, , \tag{3}$$

where $\mathbf{V}_e$ is the electron fluid velocity, $\mathbf{P}_e$ is the electron pressure tensor, $\mathbf{J}$ is the total current and $\eta$ is a resistivity. Even though we assume $m_e = 0$ in Eq. (3), we still include the electron pressure term on the right hand side. This contributes to the electric field that insures that the plasma remains quasi-neutral, and thus eliminating Debye length effects, i.e.,

$$en_e = q_i n_i \quad , \tag{4}$$

where e is the elementary charge. In Eq. (3), $\mathbf{P}_e$ is usually taken to be a scalar $\mathbf{P}_e = p_e \mathbf{1}$ but sometimes separate electron pressures along and transverse to the background magnetic field are included, as discussed later in Sec. 3. The last term in Eq. (3) is an optional resistive term. Usually $\eta$ is assumed be a scalar with a constant value. If this term is included, there is a corresponding term $(-e\eta \bullet \mathbf{J})$ in the ion equation of motion (1) to balance momentum exchange. This term is usually assumed small and is often used as a numerical means to reduce high-frequency fluctuations in the electromagnetic field, but in certain problems (e.g., magnetic reconnection), it can be physically significant, as a localized model for high frequency waves that initiate reconnection of field lines. We will discuss the pressure model and resistivity later in regard to numerical tests and specific applications.

The electric and magnetic fields in an electromagnetic hybrid model are treated in the low frequency approximation (i.e., Darwin model [6]) that ignores light waves. Thus, we use Ampere's law,

$$\nabla \times \mathbf{B} = \mu_o \mathbf{J} = \mu_o q_i n_i (\mathbf{V}_i - \mathbf{V}_e) \quad , \tag{5}$$

and Faraday's law

$$\frac{\partial \mathbf{B}}{\partial t} = -(\nabla \times \mathbf{E}) \quad . \tag{6}$$

Eq. (6) is used to advance the magnetic field in time. Eq. (5) is used to eliminate the electron flow velocity in Eq. (3), which then reduces to the equation for the electric field $\mathbf{E}$. In this low-frequency approximation, the electric field is thus not advanced in time, but instead is determined from the (updated) magnetic field, and ion density and flow velocity. Calculating the electric field at the next time step is the essential problem in all hybrid algorithms and different methods to accomplish this will be discussed throughout this chapter.

The numerical implementation of these equations in a hybrid code begins by choosing a spatial grid. Often, but not always, a rectangular grid is used. One must also decide on the location of fields and particle source terms on the grid. In the basic implementation we describe here, the electric field, ion density, ion current and electron pressure are located on the vertices of the grid, while the magnetic field is located at the center of the grid [7]. In this way, curl **E** is given correctly at the location of **B** and curl **B** at the location of **E**. Given the grid, the ion dynamics are done as in usual PIC codes: the fields are interpolated to the positions of the particle ions to give the correct local force and thus acceleration of the ions, and after the particles are moved, their contributions update the ion density and current values at the grid points. This can be done using linear, or another type of, weighting. While different hybrid algorithms may choose different ways to locate the fields and collected ion source terms on the grid, we will not discuss these methods further in this chapter. The ways that this can be done are similar to those used in PIC simulations that are described in chapter 6. However, it should be noted that some recent work has suggested that spatial resolution, particle shapes, and smoothing in hybrid codes can affect low frequency wave dispersion [8]. Thus, one always needs to apply suitable test problems to any hybrid code one is using.

The advance in time also follows PIC methodology and proceeds as follows. We start at time step N (denoted by superscript), when the particle positions $\mathbf{x}_p^N$ and the electric field $\mathbf{E}^N$ and magnetic field $\mathbf{B}^N$ are known, while the particle velocities are a half time step behind, $\mathbf{v}_p^{N-1/2}$. This allows the particle positions and velocities to "leapfrog" in time. First, the particle velocities are advanced to the next level N+1/2 with time step $\Delta t$ in the equations of motion:

$$\mathbf{v}_p^{N+1/2} = \mathbf{v}_p^{N-1/2} + \frac{q_i}{m_i}(\mathbf{E}^N + \mathbf{v}_p^N \times \mathbf{B}^N)\Delta t \quad , \tag{7}$$

And then the particle positions are similarly advanced to time level N+1

$$\mathbf{x}_p^{N+1} = \mathbf{x}_p^N + \mathbf{v}_p^{N+1/2}\Delta t \quad . \tag{8}$$

In the process the ion current is collected at time step N+1/2 and the density at N+1. We also save the old value of the ion density $n_i^N$ for later use. We note that in Eq. (7), $\mathbf{v}_p^N = \frac{1}{2}(\mathbf{v}_p^{N-1/2} + \mathbf{v}_p^{N+1/2})$, which can then be solved to get $\mathbf{v}_p^{N+1/2}$, e.g. Ref. [9].

Next, the magnetic field $\mathbf{B}^N$ is advanced to $\mathbf{B}^{N+1/2}$ using $\mathbf{E}^N$ and Faraday's law, Eq. (6):

$$\mathbf{B}^{N+1/2} = \mathbf{B}^N - \frac{\Delta t}{2}(\nabla \times \mathbf{E}^N) \quad . \tag{9}$$

By solving the electron momentum Eq. (3) for the electric field, we can evaluate $\mathbf{E}^{N+1/2}$:

$$\mathbf{E}^{N+1/2} = -\mathbf{V}_i^{N+1/2} \times \mathbf{B}^{N+1/2} - \frac{\nabla p_e^{N+1/2}}{q_i n_i^{N+1/2}} - \frac{\mathbf{B}^{N+1/2} \times (\nabla \times \mathbf{B}^{N+1/2})}{\mu_o q_i n_i^{N+1/2}} = \mathbf{E}^{N+1/2}(\mathbf{B}^{N+1/2}, \mathbf{V}_i^{N+1/2}, n_i^{N+1/2}), \tag{10}$$

because we know the ion current $\mathbf{V}_i^{N+1/2}$ from collecting the new ion particle velocities $\mathbf{v}_p^{N+1/2}$, the ion density from averaging its old and new values, $n_i^{N+1/2} = \frac{1}{2}(n_i^N + n_i^{N+1})$, and $\mathbf{B}^{N+1/2}$ from Eq. (9). On the far right side of Eq. (10) we have explicitly written out the time levels of the appropriate variables needed to compute $\mathbf{E}$, for later reference. We have also assumed an adiabatic form for the scalar electron pressure:

$$p_e = n_o T_{eo}(n_e/n_o)^\gamma \quad , \tag{11}$$

where $T_{eo}$ ($n_o$) is a reference electron temperature (density), usually with $\gamma = 5/3$ (adiabatic), so that from Eq. (4), $p_e^N$ involves just knowing $n_i^N$.

Then again, we use Faraday's Law (6), as in Eq. (9) with $\mathbf{B}^{N+1/2}$ and $\mathbf{E}^{N+1/2}$ to advance the magnetic field to $\mathbf{B}^{N+1}$. However, the advance of $\mathbf{E}^{N+1/2}$ to $\mathbf{E}^{N+1}$ is more problematic, as noted earlier. An examination of applying Eq. (10) for $\mathbf{E}^{N+1}$, shows that while $\mathbf{B}^{N+1}$ and $n_i^{N+1}$ are known, $\mathbf{V}_i^{N+1}$ is not.

Our earlier tutorial [1] described a number of ways in which this issue could be resolved. Two of those methods, namely the predictor-corrector approach [10] and the Current Advance Method and Cyclic Leapfrog (CAM-CL) approach [11], are discussed in some detail in that book chapter. Both of these methods are actively used in present day hybrid simulation codes and will be discussed later in Sec. 4. These two methods, along with a third method discussed here, will be compared later in Sec. 5.3. Moreover, a recent extension of the predictor-corrector method is presented in Sec. 5. In this updated tutorial, we restrict ourselves here to a third, and the simplest, method to advance the electric field its final half time step. In this approach, before we advance the particles to calculate $\mathbf{v}_p^{N+1/2}$ and thus accumulate the particle information on the grid to determine $\mathbf{V}_i^{N+1/2}$, we retain $\mathbf{V}_i^{N-1/2}$ (as well as $q_i n_i^N$) and then use a simple linear extrapolation to obtain:

$$\mathbf{V}_i^{N+1} = \frac{3}{2}\mathbf{V}_i^{N+1/2} - \frac{1}{2}\mathbf{V}_i^{N-1/2} \quad . \tag{12}$$

Then $\mathbf{E}^{N+1}$ can be evaluated using Eq. (10) and the update to time level N+1 is completed. A flow chart of the overall time advance process for this velocity extrapolation method is given in Fig. A1 of Karimabadi et al. [12]. Note that in this method, as well as in others discussed later [10,11], the determination of the electric field does not require a global elliptic solve, so that hybrid codes that use these methods are readily scalable to run on very large computer architectures.

In the hybrid model the shortest wavelength modes in the system are whistlers, which because of the quadratic dispersion of the wave frequency $\omega$ in terms of the wavenumber k, $\omega \sim k^2$ for parallel propagation, can be problematic. The issue can be minimized (without adding resistivity) to some degree by using a smaller time step to just advance the magnetic field from N to N+1/2, while retaining a larger time step for the ions. This sub-stepping is done using Faraday's law, Eq. (6), where $\mathbf{E}$ is expressed in terms of its source terms, Eq. (10):

$$\mathbf{B}^{N+1/2} = \mathbf{B}^N - \frac{\Delta t}{2} \nabla \times \mathbf{E}^N(\mathbf{B}^N, \mathbf{V}_i^N, n_i^N) \quad . \tag{13}$$

In this case a 4$^{th}$ order Runge-Kutta scheme [13] is used, where on the right-hand side of Eq. (13) we keep $n_i^N$ and $\mathbf{V}_i^N$ fixed and just advance $\mathbf{B}$. Specifically, we subcycle $\mathbf{B}$ from N to N+1/2 in L sub-steps so that $\Delta t' = (\Delta t/2)/L$. At the $l$-th sublevel, $l = 1,2,\ldots L$, with N' = N+$l$/L, we advance $\mathbf{B}$ to N'+1/L as follows:

$$\mathbf{B}^{N'+1/L} = \mathbf{B}^{N'} + \frac{\Delta t'}{6}\sum_{j=1}^{4} f_j \mathbf{K}_j^{N'} \quad , \tag{14}$$

where

$$\mathbf{K}_j^{N'} = -\nabla \times \mathbf{E}(\mathbf{B}^{N'} + g_j \Delta t' \mathbf{K}_{j-1}^{N'}, \mathbf{V}_i^N, n_i^N) \quad , \tag{15}$$

and $f_j = (1,2,2,1)$, $g_j = (0, \frac{1}{2}, \frac{1}{2}, 1)$ and $\mathbf{K}_0 = 0$.

This completes the discussion of a basic hybrid code algorithm that neglects electron inertia. It is the basis of the codes used in the applications discussed in the next section. This is followed in Sec. 4 with ongoing work based on well-known hybrid codes that use different hybrid algorithms and in Sec. 5 on the development of new methods that expand the range of problems that can be studied.

## 3 Examples of current hybrid code applications

Next we discuss recent applications of large-scale hybrid codes to problems of current interest in space physics. In each case the phenomena of interest can occur over significant regions of space – typically greater than several Earth radii ($R_E$). On these scales, full electron and ion dynamics cannot be modeled, except perhaps by using very small ion to electron mass ratios. On the other hand, the physical processes involve significant ion kinetic phenomena so that an MHD or Hall-MHD model is also not sufficient. Large-scale hybrid codes, either in 2D or 3D, can now capture the essential physics on scales that can be directly related to spacecraft measurements. We consider three examples: one involving phenomena upstream of planetary bow shocks, one concerning magnetic reconnection in the magnetotail, and a third illustrating more global processes centered on the Earth's magnetosheath, but extending out towards the bow shock and in towards magnetopause as well. All of the hybrid codes in this section use some form of velocity extrapolation in computing the advanced-time electric field.

## 3.1 Planetary foreshocks and bow shocks

Global hybrid simulations and multi-spacecraft observations have resulted in a dramatic increase in our knowledge of planetary ion foreshock processes and their impacts on the magnetosheath and magnetosphere/ionosphere system responsible for the formation of the bow shock [14]. These processes involve the generation of low-frequency waves driven by back-streaming ions in the foreshock as well as discontinuities in the solar wind and their interaction with the shock. The scale lengths of these structures extend from several ion inertial lengths ($d_i$) to ~ 1 $R_E$ ($R_E/d_i$ ~ 60) at Earth – scales that are easily resolved by hybrid simulations. A simulation of such processes involves the magnetized solar wind plasma flowing in one end of the simulation box toward a stationary plasma and magnetic field that represents the planet, with appropriate inflow boundary conditions on the particles and fields. Typically, the simulation runs for some period of time to set up the bow shock and foreshock self-consistently and allows the solar wind to flow through and around the shock, with corresponding outflow boundary conditions on the other sides of the simulation. We present an illustration of this technique recently applied to the solar wind interaction with Venus.

Global hybrid simulations with Alfven Mach number ($M_A$ = solar wind speed/Alfven speed) greater than three show the presence of structures in the Earth's foreshock consisting of cores of reduced density and magnetic field and rims of enhanced field and density, termed foreshock cavitons [14,15]. The size of these cavitons corresponds to ~1 $R_E$ at Earth, implying that many foreshock cavitons can fit into the Earth's foreshock [16]. The results of the 3D global hybrid simulations presented here show the formation of foreshock cavitons also occurs at Venus despite the much smaller size of its foreshock [17]. In these calculations the box size is 400 × 300 × 300 $c/\omega_{pi}$ with cell size $\Delta x = \Delta y = \Delta z = 1$ $c/\omega_{pi}$ (where $c/\omega_{pi}$ = ion inertial length = $d_i$), which for typical solar wind values is ~ 100 km. The model uses a total of $6 \times 10^8$ particles to represent the solar wind protons and ionospheric $O^+$ ions and a time step of 0.0025 $\Omega_p^{-1}$ where $\Omega_p$ is the proton gyrofrequency. Fig. 1 shows results from such a run, with panel (a) corresponding to ion temperature, log of density in panel (b), and the total magnetic field strength in the bottom panel. In these plots X is the direction of the solar wind flow, the magnetic field lies in the X-Y plane. Examples of foreshock cavitons are evident in the figure; their size is comparable to the radius of Venus. The figure also shows the presence of a foreshock compressional boundary (FCB) associated with enhancements in density and magnetic field at the edges of the foreshock [18,19]. As with the underlying waves forming them, foreshock cavitons are carried by the solar wind towards the bow shock. As they approach the shock, their density and magnetic field rims become larger and the core region becomes associated with highly decelerated and heated solar wind [20].

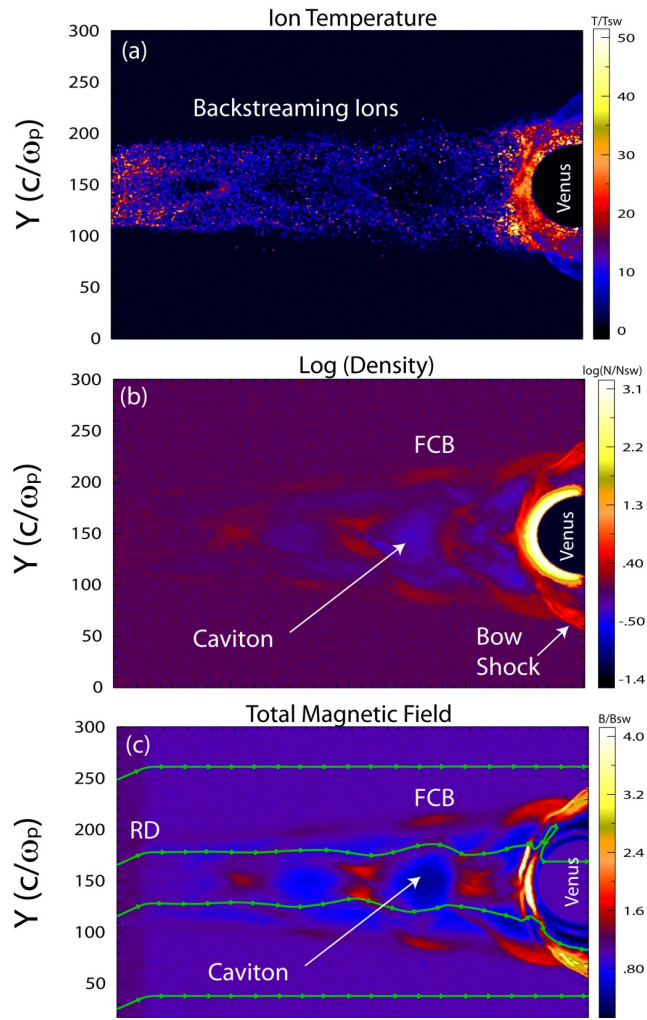

**Fig. 1.** Plots of ion temperature, density and magnetic field strength showing the structure of the Venusian foreshock. Also shown in panel (c) are magnetic field lines indicating the presence of a rotational discontinuity (RD) in the solar wind.

**3.2 Improved electron pressure closure for magnetic reconnection in the magnetosphere**

As discussed earlier, a simple isotropic, adiabatic electron pressure model is often assumed in hybrid codes. However, for applications such as for magnetic reconnection that commonly occurs at the magnetopause and in the magnetotail (and even in the magnetosheath and upstream of the bow shock, as we will see shortly), a more detailed model may be necessary to improve agreement with theory and observations. In this sub-section, we review recent results from implementing an improved electron fluid closure in hybrid simulations. The application considered—a single reconnecting current sheet—is relatively simple. Observational, theoretical, and fully kinetic numerical studies of magnetic reconnection have demonstrated that strong electron pressure anisotropy, with different effective temperatures in the directions parallel and perpendicular to the magnetic field, may develop in current sheets in space [21]. Most importantly for hybrid modeling, this electron pressure anisotropy is not confined to small electron-scale boundary layers. Rather, the electron pressure tensor may remain anisotropic over scales greater than tens of ion inertial lengths and couple back to the ion-scale system evolution.

A main mechanism for generating electron pressure anisotropy near reconnection layers is the trapping of electrons in an effective ambipolar quasi-potential $\Phi_\parallel$, which is the parallel electric field integrated along the magnetic field lines. This trapping process is well-understood in the guiding center picture, and a relatively simply model has been formulated to describe the electron velocity distribution when there is a guide magnetic field present (so that the electron particle orbits remain magnetized throughout the reconnecting current sheet) [22]. From the kinetic model, a set of anisotropic equations of state (AEoS) for the electron pressure tensor components parallel and perpendicular to the magnetic field may be derived [23]. The AEoS are similar to the Chew-Goldberger-Low (CGL) fluid closure, which gives the pressure components in terms of the density and magnetic field strength. Indeed, the AEoS reduce to the CGL scaling in the limit of strong electron trapping.

The AEoS have been implemented in reconnection models in the hybrid code H3D [24] by including the electron pressure tensor divergence in the Ohm's law for the electric field. The AEoS hybrid model has recently been validated with observations gathered by NASA's MMS mission [2] at a magnetopause reconnection site [25]. In Fig. 2, the electron pressure tensor components observed by MMS plotted here as electron density and temperatures parallel and perpendicular to the local magnetic field show remarkably good agreement with the AEoS predictions from a hybrid model set up to match the plasma parameters of this reconnection event. Note that the hybrid model with AEoS, which assume a very fast parallel transit time of electrons, reproduces this electron anisotropy even more accurately than a fully kinetic simulation performed with a reduced mass ratio of $m_i/m_e = 100$.

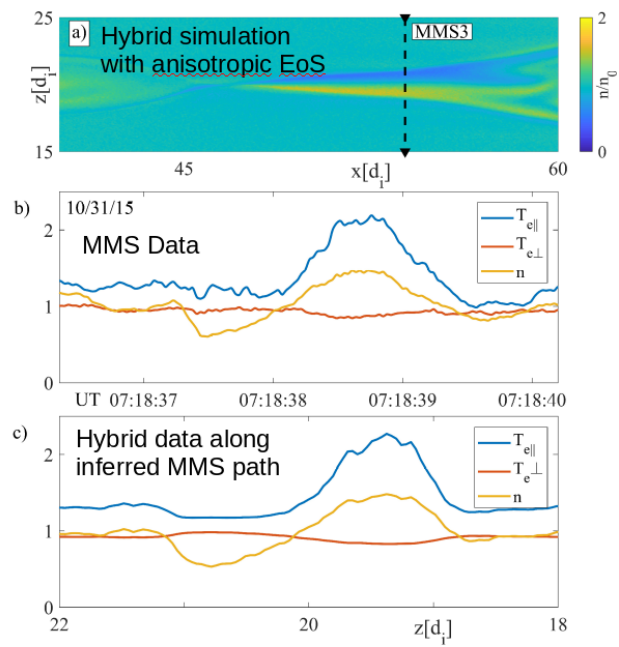

**Fig. 2.** (a) Plasma density from a hybrid simulation matching the parameters of an MMS observations of a reconnection exhaust. (b) MMS measurements of the plasma density and electron temperature components. (c) Data from the hybrid model along the cut shown in (a) displaying good agreement with the MMS observations.

The MMS event considered above contained a fairly strong guide magnetic field $B_g/B_0 > 1$. The effect of the electron pressure anisotropy is more pronounced in other regimes [26, 27] with weaker guide magnetic fields, when the electron beta is high near the X-line and the electron pressure becomes more dynamically important. For upstream electron betas of a few percent and guide magnetic fields of $B_g/B_0 \sim 0.2 - 0.6$, a regime exists where the electron pressure is close to the firehose threshold, $p_\parallel - p_\perp \sim B^2/\mu_o$, throughout the exhaust region, meaning the electron pressure anisotropy nearly balances the magnetic tension forces on the electron fluid. This allows strong perpendicular electron currents to flow, forming elongated current sheets extending from the X-line. These extended electron current sheets, as illustrated in Fig. 3, reach across ion scales and are limited in the simulations only by the size of the numerical domain.

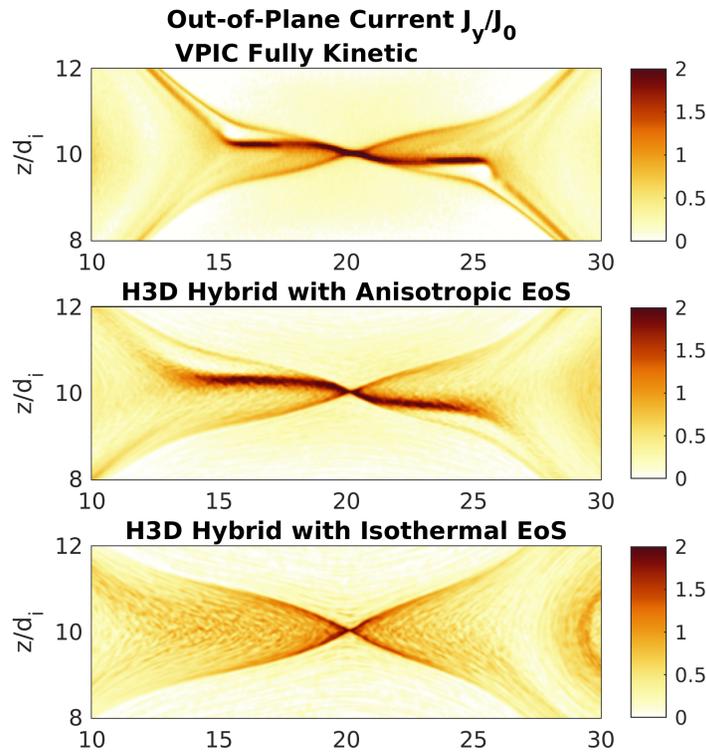

**Fig. 3.** Out-of-plane current density in a fully-developed reconnection region from (a) a fully kinetic simulation, (b) a hybrid simulation including electron pressure anisotropy, and (c) a hybrid simulation with isothermal electron closure. The electron pressure anisotropy supports the elongated current layer in panels (a) and (b).

**3.3 Magnetosheath: Effects of the bow shock, turbulence and reconnection**

Due to the availability of *in situ* spacecraft measurements, the Earth's bow shock provides an important example for basic studies of fast magnetosonic shocks and turbulence. While observations provide a local snapshot of the plasma, simulations provide a larger scale as well as dynamical view of the environment. This complimentary approach has proven quite effective in uncovering fundamental physics of collisionless shocks and associated turbulence. In this sub-section we discuss global hybrid simulations of the solar wind interacting with the Earth's magnetosphere, as described in detail in Karimabadi et al. [28]. Details of the hybrid code used in these calculations are found in Karimabadi et al. [29]. While results from the simulations displayed as individual figures give the reader some sense of the complex interactions that are occurring, computer-generated movies of the simulations discussed in this sub-section are also available with links given in the text.

In these 2D simulations, as with those reported earlier in Sec. 3.1, the bow shock, as well as the foreshock region upstream and the magnetosheath region downstream, are formed self-consistently in the simulation. The solar wind is uniformly and continuously injected from the left boundary and interacts on the right with the planetary magnetic field, a line dipole. The simulation is run for a long period of time in order to remove initial transients. Then after some period of time, a rotational discontinuity is introduced at the left boundary by changing the direction of the interplanetary magnetic field.

While Karimabadi et al. [28] considered the properties of magnetosphere under a variety of conditions, the focus here is on their resolution to a key issue that was not previously reconciled between observations and simulations. This issue was related to observations of a small reconnection event in the magnetosheath by Retino et al. [30], which they attributed to the effect of turbulence. It remained unclear whether this was just an isolated, anomalous event or a common feature of the magnetosheath turbulence. And it remained a mystery as to why no kinetic simulation of the quasi-parallel shocks, either in isolation or in a global kinetic simulation of the magnetosphere, had seen evidence for such reconnection events. Karimabadi et al. [28], using a series of peta-scale hybrid simulations and advanced visualization techniques, were able to resolve these questions. In the first simulation shown here, the system size is 8192 x 8192 ion inertial lengths ($d_i$) with cell size of 1.0 $d_i$ and 200 particles per cell. Fig. 4a shows an intensity plot of the ion density of a segment of the simulation that clearly demonstrates the formation of waves in the foreshock and a portion of the magnetosheath. In the magnetosheath, current sheets and magnetic islands can also be seen. The magnetic islands are a consequence of reconnection events, first reported by Retino et al. [30], which serve as an important dissipation mechanism.

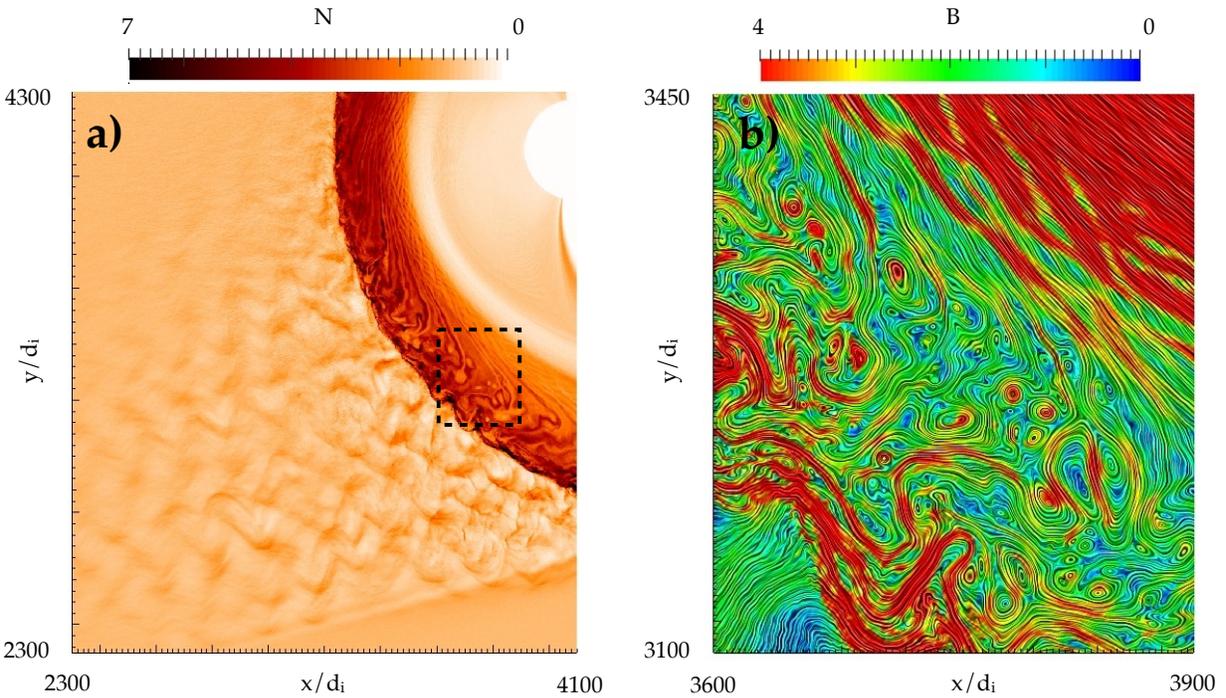

**Fig. 4.** Foreshock and magnetosheath turbulence in a 2D hybrid simulation describing interaction of solar wind injected from the left boundary with a dipolar field. Left: ion density; Right: LIC visualization of the magnetic field structure in a sub-region marked on the left panel. The scales are normalized to the ion inertial length $d_i$.

Fig. 4b shows a close up of the quasi-parallel magnetosheath. The Line Integral Convolution (LIC) technique, which is a unique method to represent streamlines of a vector field that is described in the appendix of Karimabadi et al. [28], is used to illustrate the structure of the magnetic field B with the color corresponding to the magnitude of B. Many magnetic islands are observed near the shock surface and filling the magnetosheath to the vicinity of the magnetopause. The dynamical evolution of the magnetosphere in this case as well as a zoomed in area of the simulation around the quasi-parallel magnetosheath are shown in two movies [link to ESM: S1_den_large.mp4] and [link to ESM: S1_den_zoomed.mp4], respectively. These findings demonstrated that small-scale reconnection events are a common byproduct of the turbulence developed downstream of quasi-parallel shocks, thus establishing a link between shocks, turbulence and reconnection. Subsequent high-resolution spacecraft observations with MMS have confirmed the presence of reconnection events in the magnetosheath and have strengthened their association with turbulence , e.g. Refs. [31,32].

## 4 Other hybrid algorithms, codes, and applications

In the previous section, we showed examples of the use of 2D and 3D hybrid codes applied to recent problems in magnetospheric physics. All of these calculations use algorithms based on the extrapolation of the ion velocity in order to advance the electric field to the next time step. In this section, we first give brief examples of other hybrid codes based on different algorithms that use alternative methods to advance the electric field and how these codes are being applied. We include a brief description of hybrid modes that include finite electron inertia. We then discuss in a bit more detail several new algorithms that are designed to deal more carefully with issues related to energy and momentum conservation. We conclude this section with comments related to comparison of different hybrid algorithms.

The solar wind is a very dynamic plasma environment, especially in the kinetic ion regime for which hybrid codes provide optimal capabilities. Hybrid codes used by Liewer et al. [33], by Hellinger et al. [34-37], and by Franci et al. [38-40] all employ the CAM-CL algorithm developed by Matthews [11]. It contains two significant changes in the calculation of the electromagnetic fields relative to the hybrid algorithm presented in Sec. 2. First, the advance from time step N to time step N+1 again involves a slight approximation. In the preceding discussion related to Eq. (12), an extrapolation to advance $\mathbf{V}_i^{N+1/2}$ to $\mathbf{V}_i^{N+1}$ was used in order to advance the electric field to the next time step, $\mathbf{E}^{N+1}$. Instead, in the CAM-CL approach the equation for the ion velocity moment is used:

$$\frac{\partial \mathbf{V}_i}{\partial t} = \frac{q_i}{m_i}(\mathbf{E} + \mathbf{V}_i \times \mathbf{B}) - \frac{1}{n_i m_i} \nabla \cdot \mathbf{P}_i \quad . \tag{16}$$

The right hand side of this equation contains the next higher velocity moment, namely the ion pressure tensor $\mathbf{P}_i$, which is not needed (or calculated) in a hybrid algorithm. Instead, $\mathbf{V}_i^{N+1}$ is approximated by ignoring this term, i.e., by keeping only the "free-streaming" contribution: $(q_i/m_i)(\mathbf{E}^{N+1/2} + \mathbf{V}_i^{N+1/2} \times \mathbf{B}^{N+1/2})$. The second change in this algorithm is to keep two copies of the magnetic field, one associated with integer time steps (N, N+1,…), the other with half times steps (N+1/2, N+3/2,…) that are leapfrogged over each other. As before, each version of $\mathbf{B}$ is advanced using Faraday's Law with smaller substeps. These two differences, i.e., using a moment approach to update the current and the leap-frogged magnetic field advance, explains why this method is referred to as Current Advance Method and Cyclic Leapfrog (CAM-CL). Matthews [11] provides a flow diagram (his Fig. 1) to better explain his algorithm.

To study the development of low frequency instabilities in the solar wind, a moving box technique developed by Liewer et al. [33] was used by Hellinger et al. [34] to follow a parcel of solar wind as it flows out from the sun. As a consequence of the expansion of the solar wind, the background magnetic field and density fall off with radial distance. In the simulations this effect is included as an extra force on the ions that leads to a cooling of the expanding plasma and the development of ion temperature anisotropies. Using this technique, Hellinger and Trávnícek [35] have studied the growth and nonlinear development of parallel and oblique firehose instabilities showing how these two modes effectively limit the ion temperature anisotropy. In the presence of large scale Alfvenic turbulence characteristic of the solar wind, the development and long-time behavior of both firehose ($T_{i\|} > T_{i\perp}$) and mirror ($T_{i\perp} > T_{i\|}$) instabilities, and their overall effects on the wave turbulence and the ion temperature, combine in a rather complex manner [36,37].

The CAM-CL algorithm has also been used by Franci et al. [38-40] in their 2D/3D CAMELIA hybrid code to study the development and evolution of plasma turbulence in the solar wind. Here, as opposed to outward expanding plasma in Hellinger et al. [36], the plasma is at rest and effects associated with the expansion of the solar wind are neglected. In this case the turbulence is applied as long wavelength magnetic field fluctuations at t = 0, and then the system evolves to shorter wavelengths in time, eventually reaching a quasi-steady state. The power spectrum of the perpendicular magnetic fluctuations, $P(k_\perp)$, in this stage varies at long wavelengths as $k_\perp^{-5/3}$, consistent with fluid turbulence. In regions corresponding to ion scales (i.e., ion inertial lengths or ion gyroradii), the power spectrum changes and varies as $k_\perp^{-3}$, characteristic of kinetic Alfvenic wave turbulence. The 2D simulations [39] are also used to examine how the number of simulation particles per cell and the size of the applied (constant) resistivity change the results and indicate where in parameter space the results are determined by physical properties and not by numerical quantities. With these effects having been quantified, 3D simulations [40] can then be done and compared with the 2D results, using similar initial conditions. It is found that similar spectral indices occur in perpendicular magnetic fluctuations, but smaller indices are found for density and velocity fluctuations in 3D, suggesting the importance of compressional effects.

Other hybrid codes are in common use for a wide variety of applications. For example, Brecht et al. [41] have carried out 2D and 3D simulations of the solar wind interaction with Mars and also Venus. Their code HALFSHEL uses the predictor-corrector algorithm of Harned [10], which will be discussed later in this section. The solar wind interaction with Mars, for example, is complicated by the eccentricity of the Mars orbit, the variability of the solar wind, and the presence of crustal magnetic fields, which make for challenging interpretations of observations from the MAVEN spacecraft [42]. Very good comparisons between the 3D hybrid simulations and MAVEN data concerning the ion loss rate from the planet have been carried out by Ledvina et al. [43]. Another example of a solar wind interaction is with the surface of the moon, as studied by Lipatov et al. [44] using 3D hybrid simulations. In this case various ion species (e.g., $He^+$, $He^{++}$, $Na^+$) are excited, escape from the moon's surface, and are picked up by the solar wind. The boundary conditions that model the conducting surface of the moon are also included in the model. The properties of the various ion species that are picked up by the solar wind have been compared with fly-by observations of the moon from the ARTEMIS spacecraft [45]. Continuing in the planetary vein, the work by Trávnícek et al. [46] regarding the interaction of the solar wind with Mercury should also be acknowledged, especially in regard to the excellent interpretation of the data from flybys and then orbits of the planet by the Mercury MESSENGER spacecraft [47]. For example, their hybrid simulations (using the CAM-CL method) are able to explain the differences in waves seen upstream and downstream of the bow shock and the absorption or escape of solar wind ions from the planet. Other simulations have investigated the differences in the interaction with Mercury's magnetosphere with the solar wind for both northward and southward interplanetary magnetic field [48].

The CAM-CL algorithm has also been used in 2D and 3D hybrid simulations to investigate in detail the structure of high Alfven Mach number ($M_A > 3$), perpendicular (upstream magnetic field perpendicular to the shock normal), collisionless shocks by Burgess et al. [49]. Their very finely resolved simulations have studied in detail the ion-driven waves in the ramp region just upstream of the shock ramp. In particular, the inclusion of the third spatial dimension allows fluctuations propagating parallel to the magnetic field to interact with ions reflected by the shock to create a complex pattern of obliquely propagating waves. Shock waves caused by coronal mass ejecta overtaking a slow portion of the solar wind have been studied by Gargaté et al. [50], using their 2D/3D hybrid code dHybrid [51], and subsequently employed by Caprioli and Spitkovsky [52] to model astrophysical shock waves. The structure of shock waves has also recently been studied with a new hybrid code similar to dHybrid, developed by Vshivkova et al. [53]. Over the years Giacalone and colleagues have used hybrid codes (based on velocity extrapolation) to investigate particle acceleration at shocks. An example of their recent work is a 3D simulation study of thermal proton acceleration at parallel shocks [54].

The interaction of the solar wind with the Earth's magnetosphere has been studied for a number of years by Lin et al., [55], based on an algorithm originally developed by Swift [56]. This code is unique in that it uses a non-rectangular grid that conforms better to the shape of the bow shock and the magnetopause. The field solver uses a combination of extrapolation and predictor-corrector techniques. This code has been recently used to understand the generation of short wavelength kinetic Alfven waves in the high latitude magnetotail due to the excitation of shear Alfven waves by fast Earthward flows during substorms [57].

Finally, while we have emphasized hybrid models in which electron inertia is neglected, more general hybrid models that retain electron inertia should also be mentioned. Much interest and development has occurred in this areas, especially in recent years, motivated by new space observations (e.g., MMS) and laboratory experiments of magnetic reconnection. An early type of finite-mass hybrid code for laboratory plasmas was developed decades ago [58], and more recently an algorithm based on a reduced set of electromagnetic field equations [59] was used to simulate small-scale structures associated with the electron diffusion region in magnetic reconnection. This type of reconnection physics is now mostly investigated with very large-scale full particle codes, e.g., Ref. [60]. Discussion of issues unique to these types of hybrid algorithms are found in the monograph by Lipatov [61] and summarized by Winske et al. [1]. Recently, interest in finite-electron mass hybrid algorithms has resurfaced due to the development of new fluid electron models included in a generalized quasi-neutral model by Amano [62] and in a more rigorous method for including electron inertia by Muñoz et al. [63], which is the basis of the CHIEF code. These models are described in detail in chapters 9 and 10 in this book. In addition, there is continued interest in Vlasov codes (Chapter 5) and implicit PIC codes, e.g., [64,65] and chapter 8.

## 5 New hybrid algorithms

While most of the discussion in this section thus far has addressed relatively recent results obtained with hybrid algorithms that are based on methods that have been around for decades, there have been recent and important developments in new algorithms. These new efforts are driven by the need to improve the accuracy of hybrid calculations that are now able to run on many thousands of processors with billions (and even trillions) of particles for on the order of a million time steps. In such very large, time-consuming, and expensive calculations, the issues of energy conservation and slow-growing numerical instabilities become more important. It requires a more careful examination of how the particle information is transferred to the grid and how the electromagnetic fields are solved, with the ever-present constraint of hybrid models that the electric field is not advanced in time. Here we discuss two very different approaches to address these questions. In addition, the discussion of a third, relatively new code, HYPERS [66], developed by Omelchenko and Karimabadi that includes asynchronous time advancement, is considered in more detail in Chapter 13.

## 5.1 An advanced hybrid model

One of the new hybrid codes, PEGASUS, has been developed by Kunz et al. [67] specifically for astrophysical applications. The code is unique in that additional force terms (i.e., gravity, Coriolis and centrifugal) have been introduced to model rotating and shearing plasmas that occur in astrophysics, using a shearing sheet formulation.

The code also uses a refined predictor-predictor corrector algorithm to more accurately advance the electric and magnetic fields to the next time step. To explain this procedure, it is useful to recall the traditional predictor-corrector scheme used in hybrid codes [10] in order to obtain the advanced electric field $\mathbf{E}^{N+1}$. As shown earlier, Eqs. (1-10), the ion velocities and the magnetic field can be advanced to time level N+1/2, and the ion density can also be determined at that time, which means that $\mathbf{E}^{N+1/2}$ can also then be calculated. A predicted value for the electric field $\mathbf{E}'^{N+1}$ can be readily obtained from the time-centered equation of the form,

$$\mathbf{E}^{N+1/2} = \frac{1}{2}(\mathbf{E}^N + \mathbf{E}'^{N+1}) \qquad . \tag{17}$$

by rearranging:

$$\mathbf{E}'^{N+1} = 2\mathbf{E}^{N+1/2} - \mathbf{E}^N \qquad . \tag{18}$$

Using this $\mathbf{E}'^{N+1}$ in Faraday's Law (6), a predicted magnetic field $\mathbf{B}'^{N+1}$ is then obtained from the known field $\mathbf{B}^{N+1/2}$. The particles are then advanced ahead in these predicted fields an additional time step in order to collect the moments, $n_i'^{N+3/2}$ and $\mathbf{V}_i'^{N+3/2}$. Then again using Faraday's Law with $\mathbf{E}'^{N+1}$ and $\mathbf{B}'^{N+1}$, $\mathbf{B}'^{N+3/2}$ is obtained. From $\mathbf{B}'^{N+3/2}$, $n_i'^{N+3/2}$ and $\mathbf{V}_i'^{N+3/2}$, $\mathbf{E}'^{N+3/2}$ is computed as in Eq. (10), and then the corrected $\mathbf{E}^{N+1}$ can be obtained in a symmetric fashion:

$$\mathbf{E}^{N+1} = \frac{1}{2}(\mathbf{E}^{N+1/2} + \mathbf{E}'^{N+3/2}) \qquad . \tag{19}$$

With this advanced electric field and $\mathbf{B}^{N+1/2}$, $\mathbf{B}^{N+1}$ is then computed using Faraday's Law. The flow chart in Fig. A1 of Karimabadi et al. [12] nicely summarizes the overall algorithm.

In the predictor-predictor corrector scheme of Kunz et al. [68], instead of using Eq. (18), one looks for a more time-centered approach so that the desired predicted quantity appears on the left side of Eq. (17), while the quantities on the right-hand side are either known or previously predicted. Thus, they would consider the predicted electric field of the form (18) as an unacceptable starting point. In their formulation, $\mathbf{v}_p^N$ and $\mathbf{x}_p^N$ are known at the same time, so the equations look a bit different, but the idea is similar. An overview of the process described here is shown schematically in their Fig. 2. In this case, a predicted $\mathbf{B}'^{N+1}$ is obtained from Faraday's law with $\mathbf{E}^N$. A predicted $\mathbf{E}'^{N+1}$ is computed from $\mathbf{B}'^{N+1}$, $n_i^N$ and $\mathbf{V}_i^N$ (i.e., computed using Eq. (10), but with quantities at mixed time levels) and then $\mathbf{E}'^{N+1/2}$ is obtained in a time-centered manner:

$$\mathbf{E}'^{N+1/2} = \frac{1}{2}(\mathbf{E}^N + \mathbf{E}'^{N+1}) \qquad , \tag{20}$$

and $\mathbf{B}'^{N+1/2}$ is determined from $\mathbf{B}^N$ and $\mathbf{B}'^{N+1}$ in the same manner as (20).

The particles are then advanced to time level N+1 using $\mathbf{E}'^{N+1/2}$ and $\mathbf{B}'^{N+1/2}$ to obtain the estimated moments $n_i'^{N+1}$ and $\mathbf{V}_i'^{N+1}$, which along with $\mathbf{B}'^{N+1}$, allows computing a second predictor electric field $\mathbf{E}''^{N+1}$, using Eq. (10). In addition, from $\mathbf{E}'^{N+1/2}$, $\mathbf{B}'^{N+1/2}$ and Faraday's Law, $\mathbf{B}''^{N+1}$ is computed. Then again, in the time-centered form, one obtains a second predicted electric field at time level N+1/2:

$$\mathbf{E}''^{N+1/2} = \frac{1}{2}(\mathbf{E}^N + \mathbf{E}''^{N+1}) \qquad ; \tag{21}$$

and likewise $\mathbf{B}''^{N+1/2}$ is obtained from $\mathbf{B}^N$ and $\mathbf{B}''^{N+1}$. $\mathbf{E}''^{N+1/2}$ and $\mathbf{B}''^{N+1/2}$ are then used to push the particles ahead to time level N+1, obtaining $\mathbf{v}_p^{N+1}$ and $\mathbf{x}_p^{N+1}$. In addition, the moments $n_i^{N+1}$ and $\mathbf{V}_i^{N+1}$ are obtained, and also $\mathbf{B}^{N+1}$ that is computed from $\mathbf{B}''^{N+1/2}$ and Faraday's Law with $\mathbf{E}''^{N+1/2}$. Then finally $\mathbf{E}^{N+1}$ can be computed since $n_i^{N+1}$, $\mathbf{V}_i^{N+1}$, and $\mathbf{B}^{N+1}$ are all known.

Kunz et al. [67] have carried out a number of numerical tests to validate their new code, including Landau damping, linear Alfven and whistler waves, parallel shocks, and a magneto-rotational instability. PEGASUS has also been used to study the excitation and nonlinear evolution of the firehose ($p_{i\|} > p_{i\perp}$) and mirror ($p_{i\perp} > p_{i\|}$) instabilities ($p_{i\|,\perp}$ = ion pressure, along or transverse to the magnetic field) in a plasma with a persistent linear shear, such as occurs in the solar wind or galactic accretion flows [68]. In the case where the magnetic field magnitude decreases so that $p_{i\perp}$ decreases, the firehose instability is excited and saturates at a low level, consistent with the stability threshold, due to ion scattering. In the opposite case of an increasing magnetic field and thus increasing $p_{i\perp}$, the developing ion anisotropy drives the mirror instability to large amplitudes. Saturation in this case is due to ion trapping in the large fluctuations produced by the instability.

**5.2 An implicit hybrid model**

A second new hybrid approach has been investigated by Stanier et al. [69]. This algorithm is unique here because, unlike all of the previous hybrid codes we have discussed in this chapter, it is fully implicit and is designed so that energy and momentum can be conserved to any desired degree of accuracy. (Other hybrid codes do not conserve these due to spatial or temporal truncation errors.) The work builds on previous implicit PIC methods [70, 71] that give energy (but not momentum) conservation. As in other hybrid codes, the magnetic field is advanced in time using Faraday's law, and the electric field is evaluated from the electron momentum equation with $m_e = 0$, i.e., as in Eq. (10). It should be noted that in their paper Stanier et al. prefer to use the vector potential **A** rather than the magnetic field **B,** and they solve a separate equation for the electron pressure that contains terms for resistive heating and electron heat flux. For simplicity of our presentation we will continue here to use **B**, and assume $p_e = 0$ as well as neglect resistivity. And in the spirit of the previous discussion of other hybrid codes, we will also concentrate on the temporal advance of the electric and magnetic fields.

Stanier et al. [69] assume at time level N, the particle positions $\mathbf{x}_p^N$ and velocities $\mathbf{v}_p^N$, as well as the magnetic field $\mathbf{B}^N$ are known, while the electric field is known at the previous half time step, $\mathbf{E}^{N-1/2}$. To advance the electric field to the next time level (always the problem in hybrid codes!), one assumes a guess for the advanced electric field, $\mathbf{E}'^{N+1/2}$. (How this can be done will be discussed later.) With this assumed electric field, the magnetic field can be advanced to time level N+1 using Faraday's Law to obtain $\mathbf{B}'^{N+1}$ and an equation of the form (20) can be used to compute $\mathbf{B}'^{N+1/2}$. With these fields at time N+1/2, the particles are moved a half-time step ($\Delta t/2$) to $\mathbf{x}_p'^{N+1/2}$ and $\mathbf{v}_p'^{N+1/2}$ where the moments $n_i'^{N+1/2}$ and $\mathbf{V}_i'^{N+1/2}$ are gathered at the same time level. At this stage, the initial guess of $\mathbf{E}'^{N+1/2}$ is compared with the right-hand side of Ohm's law using Eq. (10) (and setting $p_e = 0$ here). The difference between the two (**F**) values is called the 'residual' and given by

$$\mathbf{F}(\mathbf{E}^{N+1/2}) = \mathbf{E}'^{N+1/2} + \mathbf{V}_i'^{N+1/2} \times \mathbf{B}'^{N+1/2} + \frac{\mathbf{B}'^{N+1/2} \times (\nabla \times \mathbf{B}'^{N+1/2})}{\mu_o q_i n_i'^{N+1/2}} \quad . \quad (22)$$

The initial guess $\mathbf{E}'^{N+1/2}$ is therefore exact when $\mathbf{F}(\mathbf{E}^{N+1/2}) = 0$, but unfortunately this is extremely unlikely! Instead, the error is calculated as the Euclidean norm, $|\mathbf{F}|$, and compared to a user specified tolerance, $\varepsilon$. If $|\mathbf{F}| > \varepsilon$, the whole process is repeated with a new guess $\mathbf{E}''^{N+1/2}$ (and potentially $\mathbf{E}'''^{N+1/2}$ and so on) until the condition $|\mathbf{F}| < \varepsilon$ is met and the iteration is stopped. At this point, with $\mathbf{E}'^{N+1/2}$ known, $\mathbf{B}^{N+1}$ is calculated (as well as $\mathbf{B}^{N+1/2}$), and particle positions $\mathbf{x}_p^{N+1}$ and $\mathbf{v}_p^{N+1}$ are calculated for the next time step using the full time step ($\Delta t$). It is found in Stanier et al. [69] that the momentum and energy conservation errors are compatible to the value chosen for $\varepsilon$. Thus, they can be made to be as small as numerical round-off by choosing small $\varepsilon$, but this typically requires more iterations of the above method.

A natural question, and indeed the whole efficiency of the method, depends on the method to determine the electric field guesses. Stanier et al. [69] use a form of the well-known Newton-Krylov methods [72,73] to do this. $\mathbf{E}^{k+1}$ (the '(k+1)-th' guess of $\mathbf{E}^{N+1/2}$) is found by iteratively solving the Newton-linearized equations,

$$\frac{\partial \mathbf{F}}{\partial \mathbf{E}}\Big|^k (\mathbf{E}^{k+1} - \mathbf{E}^k) = -\mathbf{F}(\mathbf{E}^k) \quad , \tag{23}$$

using the flexible-Generalized Minimum Residual method (fGMRES method [74]). Alert readers of this article would note that an initial guess with k = 0 is still needed at each time step. The form of this guess is not so critical due to the iterative nature of the algorithm, and it can be taken to be the electric field at the previous time step (extrapolation).

As well as using implicit techniques to solve for the electric field, Stanier et al. [69] also advance the particles with an adaptive and implicit sub-cycling method (in contrast to sub-cycling the fields in the earlier described explicit codes), and use an orbit-averaging technique to reduce the numerical noise (also used previously in an [electrostatic] hybrid code by Sturdevant et al. [75]). The details are probably beyond the concern of most readers of this chapter – the interested reader is referred to the references in Stanier et al. [69]. We do note that sub-cycling the ions is particularly suited for magnetospheric simulations, where ions in the strong dipole field of the Earth have larger gyro-frequencies (and need smaller time steps) than those in the interplanetary magnetic field.

The utility of this method has been analyzed in a number of test problems in Stanier et al. [69]. In addition to several electrostatic problems, the well-known one-dimensional electromagnetic ion cyclotron instability driven by a temperature anisotropy ($T_\perp/T_\parallel > 1$) (e.g., [76]) that generates the most unstable waves parallel to the background magnetic field and the equally popular GEM reconnection challenge problem [77] have been successfully simulated. Moreover, excellent results have been obtained for the cold ion beam propagation problem, which has been a persistent issue in hybrid codes for some time [78].

In the long run, how well an implicit hybrid code will perform on more complex problems remains to be demonstrated. The strength of implicit methods, with the possibility of using larger time steps, having better control of energy conservation, and suppressing the impact of short-wavelength whistler waves, makes this technique well worth more study. For hybrid codes, and indeed for many other computational plasma physics techniques as well, there is always a trade-off between efficiency, stability and overall energy conservation issues.

**5.3 A brief note on comparison of hybrid algorithms**

Having discussed a number of different hybrid algorithms that have been implemented in various production codes used by a number of researchers on a wide range of problems, we conclude this section with a brief discussion of test problems, numerical checks and comparisons with different algorithms (since there will always be some differences in the results, as discussed earlier [8]). This discussion is based on our previous tutorial [1] and comparison of results from different hybrid algorithms in the Appendix of Karimabadi et al. [12]

Our earlier tutorial [1] employs two-dimensional simulations based on either the velocity-extrapolation algorithm discussed in Sec. 2 or a predictor-corrector formulation described in that article as well as earlier in this section to compare and contrast results of two equal-density, field-aligned counter-streaming ($\pm V_o = v_A$ = Alfven speed) ion beams that generate Alfven waves. Qualitative comparisons of results from runs made with these two different algorithms are described (in their Figs. 1,2), along with examining the effects of smoothing (in their Fig. 3) and changing the resistivity (in their Fig. 4) on the spectrum of high-frequency whistlers, ion heating and energy conservation. The effects on the results obtained in both codes are most noticeable at short wavelengths.

Karimabadi et al. [12] carried out the most detailed, and quantitative comparison of the velocity extrapolation, predictor-corrector and the CAM-CL algorithms in two-dimensional test problems involving a moving discontinuity of two plasma states with differing temperatures and densities and parallel or anti-parallel magnetic fields in the two adjacent regions. The temperatures of the cold ions in each case are compared in the three algorithms and with different size time steps in their Tables A1 and A2. A quick rule of thumb is that all three algorithms give "excellent" results for most practical situations, but in problems where more precision in the wave properties at short wavelength or better energy conservation are needed, not surprisingly, the more accurate, predictor-corrector algorithm performs better. The presence of short-wavelength whistlers can be reduced by sub-cycling both the electric and magnetic field solves. Some adjustments to these basic algorithms that involve when the ion charge is collected (i.e., at time level N+1 at the new particle positions or at time level N+1/2 when the velocities have been advanced) that can improve performance in certain problems, such as in cold ion beam propagation [12].

## 6 The future of hybrid codes

We conclude with a brief look into the future using hybrid codes to investigate space plasma phenomena, focusing on three issues. First, there is expected to be continued major community interest in modeling and understanding plasma physics phenomena and their relation to spacecraft observations. This will naturally involve vigorous investigation of processes occurring in the near-Earth environment, e.g., the Earth's bow shock, foreshock, magnetosheath, magnetopause, magnetotail and observations from MMS and other spacecraft. There will also be sustained interest in phenomena observed at the moon and planets, such as Mercury, Venus, Mars, Jupiter, Saturn and the other outer planets as well. Processes occurring at the sun and also in the solar wind, as well as at objects residing in the wind, such as comets and asteroids, will also remain of significant importance, especially with the launch of Parker Solar Probe and Solar Orbiter. In all of these situations, as we have seen throughout this chapter, hybrid calculations will play a unique and important role in the interpretation of observations, direct comparison with measurements, and in understanding integrated physical processes (as demonstrated in Sec. 3.3). Overall, these interests will remain the principal driver for hybrid simulations in the foreseeable future.

Second, there will be continuing improvements in algorithms, not only directly for hybrid codes but for plasma modeling in general. Specifically for hybrid algorithms, one expects continuing efforts to find better ways to reduce effects associated with short-wavelength whistler waves, dynamic methods to improve computational efficiency by varying time-steps and/or cell-sizes during the simulation, In addition, in recent years there has been further development of finite-electron mass hybrid codes, quasi-neutral models, implicit methods, and Vlasov codes (see Chapters 9, 10, 8, 5, respectively), and even renewed interest in localized embedding of kinetic electron physics [1]. One expects that such code improvements will remain interesting and challenging problems, and perhaps important and compelling as well.

Finally, there is ongoing and significant continuing development of new computer architectures (e.g., exa-scale computing), visualization techniques, asynchronous time-stepping methods (Chapter 13), and diagnostics in all areas of computational physics; all of these areas are changing very rapidly. Major drivers in these areas are likely to occur in areas other than space physics, where more abundant resources are available to bear on their problems. But, as happened frequently in the past, space plasma physics will likely benefit greatly from these advances in other fields.

In rewriting the 2003 version of this tutorial, it was most interesting to view how computational space plasma physics, even in the narrow perspective of hybrid codes, has on the one hand remained fairly steady, and yet at the same time, has shown significant changes. The "grand challenge" presented to graduates of the International Space Simulation School 20 years ago was the issue of embedding kinetic electron physics in hybrid codes. However, the development of very large scale PIC codes and the astonishing detailed measurements from MMS in recent years have rendered this challenge to be of secondary importance. It has now been replaced with the greater challenge of using hybrid codes (3D, of course, but also 2D and even 1D where appropriate) to analyze and interpret the complex integrated plasma physics that occurs at all boundary layers in space.